\documentclass[preprint,prc,aps,showpacs,showkeys,groupedaddress,floatfix]{revtex4}
\usepackage{epsfig}
\usepackage{dcolumn}
\usepackage{bm}
\usepackage[latin5]{inputenc}
\usepackage{graphics}
\usepackage{graphicx}
\usepackage{epsfig}
\usepackage{amssymb}
\usepackage{amsmath}
\begin{document}
\title{Analytical Solutions to the Hulth\'{e}n and the Morse Potentials by using the Asymptotic
Iteration Method}
\author{O. Bayrak\dag\ddag \quad and I. Boztosun\ddag}
\affiliation{\dag Faculty of Arts and Sciences, Department of
Physics, Bozok University, Yozgat, Turkey} \affiliation {\ddag
Faculty of Arts and Sciences, Department of Physics, Erciyes
University, Kayseri, Turkey}

\date{\today}
\begin{abstract}
We present the exact analytical solution of the radial
Schr\"{o}dinger equation for the deformed Hulth\'{e}n and the Morse
potentials within the framework of the Asymptotic Iteration Method.
The bound state energy eigenvalues and corresponding wave functions
are obtained explicitly. Our results are in excellent agreement with the findings of
the other methods.
\end{abstract}

 \maketitle

\section{Introduction}
In recent years, the energy eigenvalues and corresponding
eigenfunctions between interaction systems have raised a great deal
of interest in relativistic quantum mechanics as well as in
non-relativistic quantum mechanics. The exact solution of the wave
equations (relativistic or non-relativistic) is very important since
the wave function contains all the necessary information regarding
the quantum system under consideration. Analytical methods such as
the super-symmetry (SUSY) \cite{susy} and the Nikiforov-Uvarov
method (NU) \cite{Nifikorov} have been used to solve wave equations
exactly or quasi-exactly for a given potential and these potentials
are in general either the polynomial type like the Coulomb, the
harmonic oscillator \cite{fluge}  or the exponential type such as
the Eckart or Hulth\'{e}n \cite{fluge,hulthen,nu1s}, Morse
\cite{morse} or a combination of these potentials.

The Hulth\'{e}n and Morse potentials we consider in this paper have
a wide range applications. The Hulth\'{e}n potential has a very
important role in describing the bound states or continuum states
between interaction particles in the relativistic quantum mechanics
as well as the non-relativistic quantum mechanics. The Hulth\'{e}n
potential has been solved for the bound states with the NU
\cite{nu1s}, the SUSY \cite{susy1,susy2,susy3,susy4} and the shifted
$1/N$ expansion \cite{other} methods. It is also solved for the
relativistic wave equations such as Dirac \cite{nu2d, relativd1,
relativd2}, Klein-Gordon (KG) \cite{relativkg1,relativkg2} and the
Duffin-Kemmer Petiau (DKP) \cite{relativdkp1} equations.

The Morse potential has raised a great deal of interest over the years and has been one of
the most useful models to describe the interaction between two atoms in a diatomic molecule.
In this respect, Morse potential has been solved by the super-symmetry (SUSY)
\cite{susy}, the Nikiforov-Uvarov method (NU)\cite{Nifikorov}, the hypervirial perturbation
method (HV)\cite{killingbeck}, the shifted and modified shifted 1/N expansion methods \cite{bag}
as well as the variational method \cite{variational}.

Since these potentials have been extensively used to describe the
bound and the continuum states of the interactions systems, it would
be interesting and important to solve the non-relativistic radial
Schr\"{o}dinger equation for the Hulth\'{e}n and Morse potentials.
Recently, an alternative method called as the Asymptotic Iteration
Method (AIM) for solving second-order homogeneous linear
differential equation has been developed by \c{C}ift\c{c}i \emph{et
al.} \cite{hakanaim1,hakanaim2} and has been applied to solve the
non-relativistic radial Schr\"{o}dinger equation or the relativistic
Dirac equation.

In this paper, our aim is to solve the deformed Hulth\'{e}n
potential and the Morse potentials to obtain the energy eigenvalues
and corresponding eigenfunctions within framework of the AIM.
In the next section, the asymptotic iteration method
(AIM) is introduced. Then, in section \ref{apply}, the
Schr\"{o}dinger equation is solved by the asymptotic iteration
method for the deformed Hulth\'{e}n potential and the Morse
potential. The AIM results are compared with the findings of
the other methods. Finally, section \ref{conclude} is devoted to the summary and conclusion.

\section{Overview the Asymptotic Iteration Method (AIM)}
\label{aim}
\subsection{Energy Eigenvalues}
AIM is briefly outlined here and the details can be found in
references \cite{hakanaim1,hakanaim2,bayrakJPA}. AIM is proposed to
solve the second-order differential equations of the form
\begin{equation}\label{diff}
  y''=\lambda_{0}(x)y'+s_{0}(x)y
\end{equation}
where $\lambda_{0}(x)\neq 0$ and the prime denotes the derivative
with respect to $x$. The variables, $s_{0}(x)$ and $\lambda_{0}(x)$,
are sufficiently differentiable. The differential equation
(\ref{diff}) has a general solution \cite{hakanaim1}
\begin{equation}\label{generalsolution}
  y(x)=exp \left( - \int^{x} \alpha(x_{1}) dx_{1}\right ) \left [C_{2}+C_{1}
  \int^{x}exp  \left( \int^{x_{1}} [\lambda_{0}(x_{2})+2\alpha(x_{2})] dx_{2} \right ) dx_{1} \right
  ]
\end{equation}
for sufficiently large $k$, $k>0$, if
\begin{equation}\label{termination}
\frac{s_{k}(x)}{\lambda_{k}(x)}=\frac{s_{k-1}(x)}{\lambda_{k-1}(x)}=\alpha(x),
\quad k=1,2,3,\ldots
\end{equation}
where
\begin{eqnarray}\label{iter}
  \lambda_{k}(x) & = &
  \lambda_{k-1}'(x)+s_{k-1}(x)+\lambda_{0}(x)\lambda_{k-1}(x) \quad
  \nonumber \\
s_{k}(x) & = & s_{k-1}'(x)+s_{0}(x)\lambda_{k-1}(x), \quad \quad
\quad \quad k=1,2,3,\ldots
\end{eqnarray}

Note that one can also start the recurrence relations from $k=0$
with the initial conditions $\lambda_{-1}=1$ and $s_{-1}=0$
\cite{fernandez}. For a given potential, the radial Schr\"{o}dinger
equation is converted to the form of equation (\ref{diff}). Then,
s$_{0}(x)$ and $\lambda_{0}(x)$ are determined and s$_{k}(x)$ and
$\lambda_{k}(x)$ parameters are calculated by the recurrence
relations given by equation~(\ref{iter}).

The termination condition of the method in equation
(\ref{termination}) can be arranged as

\begin{equation}\label{quantization}
  \Delta_{k}(x)=\lambda_{k}(x)s_{k-1}(x)-\lambda_{k-1}(x)s_{k}(x)=0 \quad \quad
k=1,2,3,\ldots
\end{equation}

The energy eigenvalues are obtained from the roots of the equation
(\ref{quantization}) if the problem is exactly solvable. If not, for
a specific $n$ principal quantum number, we choose a suitable $x_0$
point, determined generally as the maximum value of the asymptotic
wave function or the minimum value of the potential
\cite{hakanaim1,fernandez}, and the approximate energy eigenvalues
are obtained from the roots of this equation for sufficiently great
values of $k$ with iteration.

\subsection{Energy Eigenfunctions}
In this study, we seek the exact solution of radial Schr\"{o}dinger
equation for which the relevant second order homogenous linear
differential equation takes the following general form
\cite{hakanaim2},
\begin{equation}
\label{compare1}
 {y}''(x) = 2\left( {\frac{ax^{N + 1}}{1 - bx^{N + 2}} -
\frac{\left( {m + 1} \right)}{x}} \right){y}'(x) - \frac{wx^N}{1 -
bx^{N + 2}}y(x)
\end{equation}
the following general formula for the exact solutions for $y_n(x)$ is given by \cite{hakanaim2},
\begin{equation}\label{efson}
y_n (x) = \left( { - 1} \right)^nC_2 (N + 2)^n\left( \sigma
\right)_n { }_2F_1 ( - n,\rho + n;\sigma ;bx^{N + 2})
\end{equation}

where $(\sigma )_n $=$\frac{\Gamma ( {\sigma + n} )}{\Gamma (\sigma)
}, \quad \sigma$ = $\frac{2m + N + 3}{N + 2}$ \quad \mbox{and} \quad
$\rho$ = $\frac{( {2m + 1} )b + 2a}{( {N + 2} )b}$.

\section{Calculation of the Energy Eigenvalues and Eigenfunctions}
\label{apply} The motion of a particle with the mass $m$ is
described by the following Schr\"{o}dinger equation:

\begin{equation}
\frac{-\hbar^{2}}{2m}\left(\frac{\partial^{2}}{\partial
r^{2}}+\frac{2}{r}\frac{\partial}{\partial r}+ \frac{1}{r^{2}}
\left[\frac{1}{\sin \theta}\frac{\partial}{\partial \theta} \left(
\sin \theta \frac{\partial}{\partial \theta} \right)
+\frac{1}{\sin^{2} \theta} \frac{\partial^{2}}{\partial \phi^{2}}
\right]+V(r) \right)\Psi_{n\ell m}(r,\theta,\phi)=E\Psi_{n\ell
m}(r,\theta,\phi)\label{sch}
\end{equation}
The terms in the square brackets with the overall minus sign is the
dimensionless angular momentum squared operator, ${\bf L}^{2}$.
Defining $\Psi_{n\ell m}(r,\theta,\phi)=u_{n\ell}(r)Y_{\ell
m}(\theta,\phi)$, we obtain the radial part of the Schr\"{o}dinger
equation:
\begin{eqnarray}
\left(\frac{d^{2}}{d{r}^{2}}+\frac {2}{r}\frac{d}{dr}
\right)u_{n\ell}(r)-\frac{2m}{\hbar^{2}}\left[V(r)+\frac{\ell(\ell+1)\hbar^{2}}{2
m r^{2}} \right]u_{n\ell}(r)+\frac{2 m
E}{\hbar^{2}}u_{n\ell}(r)=0 \label{radialr}
\end{eqnarray}
It is sometimes convenient to define $u_{n\ell}(r)$ and the
effective potential as follows
\begin{equation}
u_{n\ell}(r)=\frac{R_{n\ell}(r)}{r}, \quad
V_{eff}=V(r)+\frac{\ell(\ell+1)\hbar^{2}}{2 m r^{2}}
\end{equation}
Since
\begin{equation}
\left(\frac{d^{2}}{d{r}^{2}}+\frac {2}{r}\frac{d}{dr}
\right)\frac{R_{n\ell}(r)}{r}=\frac{1}{r}\frac{d^{2}}{d{r}^{2}}R_{n\ell}(r)
\end{equation}
The radial Schr\"{o}dinger equation given by equation
(\ref{radialr}) follows that
\begin{equation}
\frac{d^{2}R_{n\ell}(r)}{d{r}^{2}}+\frac {2m}{\hbar^{2}}
\left[E-V_{eff} \right]R_{n\ell}(r)=0 \label{factorschrodinger}
\end{equation}
Instead of solving the partial differential equation (\ref{sch}) in
three variables $r$, $\theta$ and $\phi$, we now solve a
differential equation involving only the variable $r$ and the
angular momentum parameter $\ell=0$ (s-state solution).

\subsection{Deformed Hulth\'{e}n Potential Case}
The deformed Hulth\'{e}n potential \cite{hulthen} is defined as
\begin{equation}\label{dh}
  V_{DH}(r)=-Ze^{2}\delta\frac{e^{-\delta r}}{1-qe^{-\delta r}}
\end{equation}
In this formula, Z, $\delta$ and $q$ are respectively the atomic
number, the screening parameter and the deformation parameter
determining the range for the deformed Hulth\'{e}n potential. The
deformed Hulth\'{e}n potential reduces to the Hulth\'{e}n potential
form for $q=1$, to the standard Wood-Saxon potential for $q=-1$ and
to the exponential potential  for $q=0$.

Inserting the Hulth\'{e}n potential given by equation (\ref{dh})
into equation \ref{factorschrodinger} and using following
\emph{ansatzs}
\begin{equation}\label{ansatz}
-\varepsilon^{2}=\frac{2m E}{\hbar^{2}\delta^{2}}, \quad
\beta^{2}=\frac{2m e^{2}Z}{\hbar^{2}\delta}, \quad
\end{equation}
and we apply a transformation to $\delta r=x$ to the deformed Hulth\'{e}n potential.
The radial Schr\"{o}dinger equation takes the following form:
\begin{equation}\label{radyal}
 \frac{d^{2}R(x)}{dx^{2}}+\left( -\varepsilon^{2}+\beta^{2}\frac{e^{-x}}{(1-qe^{-x})} \right) R(x)=0
\end{equation}
If we rewrite equation \ref{radyal} by using a new variable of the
form $y=e^{-x}$, we obtain
\begin{equation}\label{trans}
\frac{d^{2}R(y)}{dy^{2}}+\frac{1}{y}\frac{dR(y)}{dy}
+\left[-\frac{\varepsilon^{2}}{y^{2}}+ \beta^{2}\frac{1}{y(1-qy)} \right]R(y)=0
\end{equation}
In order to solve this equation with AIM, we should
transform this equation to the form of equation \ref{diff}.
Therefore, the reasonable physical wave function we propose is as
follows
\begin{equation}\label{wave}
R(y)=y^{\varepsilon} (1-qy)f(y)
\end{equation}
If we insert this wave function into the equation \ref{trans}, we
have the second-order homogeneous linear differential equations in
the following form
\begin{equation}\label{aimschr}
\frac{d^{2}f(y)}{dy^{2}}=\left[\frac{(2\varepsilon q+3q)y-(2\varepsilon+1)}{y(1-qy)}\right]
\frac{df(y)}{dy}+\left[\frac{2\varepsilon q+q-\beta^{2}}{y(1-qy)} \right]f(y)
\end{equation}
which is now amenable to an AIM solution. By comparing this equation
with equation \ref{diff}, we can write the $\lambda_{0}(y)$ and
$s_{0}(y)$ values and by means of equation \ref{iter}, we may
calculate $\lambda_k(y)$ and $s_k(y)$. This gives:
\begin{eqnarray}\label{ini}
    \lambda_{0}&=&\left(\frac{2\varepsilon q y+3qy-2\varepsilon-1}{y(1-qy)}\right) \nonumber \\
    s_{0}&=&\left(\frac{2\varepsilon q+q-\beta^{2}}{y(1-qy)} \right)\nonumber \\
    \lambda_{1}&=&{\frac {2+{\beta}^{2}q{y}^{2}-8\,{\varepsilon}^{2}yq+4\,{\varepsilon}^{2}{y}
^{2}{q}^{2}-y{\beta}^{2}+12\,{q}^{2}{y}^{2}\varepsilon+11\,{q}^{2}{y}^{2}
-18\,\varepsilon\,yq+6\,\varepsilon-7\,qy+4\,{\varepsilon}^{2}}{{y}^{2}\left(-1+qy\right)^{2}}} \nonumber \\
   s_{1}&=& {\frac {\left (2\,\varepsilon\,q+q-{\beta}^{2}\right )\left (-2+5\,qy-2\,\varepsilon+2\,
   \varepsilon\,yq\right )}{{y}^{2}\left (-1+qy\right)^{2}}}
 \\ \ldots \emph{etc} \nonumber
\end{eqnarray}
Combining these results with the quantization condition given by
equation \ref{quantization} yields
\begin{eqnarray}
 \frac{s_0 }{\lambda _0 } & = & \frac{s_1 }{\lambda _1 }\,\,\,\,\,\, \Rightarrow
\,\,\,\,\,\,\varepsilon_{0}  =  -\frac{1}{2}\frac{q-\beta^{2}}{q} \nonumber \\
 \frac{s_1 }{\lambda _1 }  & = & \frac{s_2 }{\lambda _2 }\,\,\,\,\,\, \Rightarrow
\,\,\,\,\,\, \varepsilon_{1}=-\frac{1}{4}\frac{4q-\beta^{2}}{q} \nonumber \\
 \frac{s_2 }{\lambda _2 }  & = & \frac{s_3 }{\lambda _3 }\,\,\,\,\,\, \Rightarrow
\,\,\,\,\,\,\varepsilon_{2}=-\frac{1}{6}\frac{9q-\beta^{2}}{q} \\
\ldots \emph{etc} \nonumber
 \end{eqnarray}
When the above expressions are generalized, the eigenvalues turn out
as
\begin{equation}\label{energy}
\varepsilon_{n}=-\frac{1}{(2n+2)}\left(\frac{q(n+1)^{2}-\beta^{2}}{q}\right),\hspace{1cm} n=0,1,2,3,...
\end{equation}
Using equation \ref{ansatz}, we obtain the energy eigenvalues
E$_{n}$,
\begin{equation}\label{energyeigenvalues}
E_{n}=-\frac{\hbar^{2}}{2m}\left[\frac{\frac{m e^{2}Z}{\hbar^{2}}}{(n+1)q}-\frac{(n+1)\delta}{2}\right]^{2},
\hspace{1cm} n=0,1,2,3,...
\end{equation}
In the atomic units $(\hbar =m =e = 1)$ and for Z = 1, equation
\ref{energyeigenvalues} turns out to be
\begin{equation}\label{eigenvaluesatomic}
E_{n}=-\frac{1}{2}\left[\frac{1}{\overline{n}q}-\frac{\overline{n}\delta}{2}\right]^{2} \hspace{1cm} \overline{n}=n+1
\hspace{1cm} \overline{n}=1,2,3,...
\end{equation}
It may thus be seen that the energy eigenvalue equation is easily
obtained by using AIM. In order to test the accuracy of equation
\ref{eigenvaluesatomic}, we calculate the energy eigenvalues for
$Z=1$, $q=1$ and several values of the screening parameter. The AIM
results are compared with the Nikiforov-Uvarov \cite{nu1s} and the
shifted 1/N \cite{other} expansion methods in Table \ref{Table1}. It
may here be seen that the AIM results are in excellent agrement with
the findings of the other methods.

Now, as indicated in Section \ref{aim}, we can determine the
corresponding wave functions by using equation \ref{efson}. When we
compare equation \ref{compare1} and equation \ref{aimschr}, we find
$N=-1$, $b=q$, $a=q$, and $m=\frac{2\varepsilon-1}{2}$. Therefore,
we obtain $\rho=2\varepsilon+2$ and $\sigma=2\varepsilon+1$. Having
determined these parameters, we can easily find the eigenfunctions
$f_{n}(y)$ by using equation \ref{efson} as follows
\begin{equation}
f_{n}(y)=(-1)^{n}\frac{\Gamma(2\varepsilon_{n}+n+1)}{\Gamma(2\varepsilon_{n}+1)}
{_{2}}F_{1}(-n,2\varepsilon_{n}+n+2;2\varepsilon_{n}+1;qy)
\end{equation}
Finally, we can write the total radial wave function as below,
\begin{equation}\label{radyalwave}
R_{n}=N C_{n}y^{\varepsilon_{n}}(1-qy)_{2}F_{1}(-n,2\varepsilon_{n}+n+2;2\varepsilon_{n}+1;qy)
\end{equation}
where
$C_{n}=(-1)^{n}\frac{\Gamma(2\varepsilon_{n}+n+1)}{\Gamma(2\varepsilon_{n}+1)}$.

\subsection{Morse Potential Case}
The Morse potential is defined as
\begin{equation}\label{morse}
  V_{Morse}(r)=D_{e} \left(e^{-2\alpha x}-2e^{-\alpha x} \right)
\end{equation}
with $x=(r-r_{e})/r_{e}$ and $\alpha=a r_{e}$. Here, $D_{e}$ and
$\alpha$ denote the dissociation energy and Morse parameter,
respectively. $r_{e}$ is the equilibrium distance (bound length)
between nuclei and $a$ is a parameter to control the width of the
potential well. Inserting equation \ref{morse} into equation
\ref{factorschrodinger} and using the following \emph{ansatzs}
\begin{equation}\label{ansatzm}
-\varepsilon^{2}=\frac{2\mu r_{e}^{2}E}{\hbar^{2}},    \quad
\quad \beta^{2}=\frac{2\mu r_{e}^{2}D_{e}}{\hbar^{2}}   \quad
\end{equation}
The radial Schr\"{o}dinger equation takes the following form:
\begin{equation}\label{radyalm}
 \frac{d^{2}R_{n}(x)}{dx^{2}}+\left( -\varepsilon^{2}+2\beta^{2} e^{-\alpha x}-\beta^{2} e^{-2\alpha x}\right)R_{n}(x)=0
\end{equation}
If we rewrite equation \ref{radyalm} by using a new variable of the
form $y=e^{-\alpha x}$, we obtain
\begin{equation}\label{transm}
\frac{d^{2}R_{n}(y)}{dy^{2}}+\frac{1}{y}\frac{dR_{n}(y)}{dy}
+\left[-\frac{\varepsilon^{2}}{\alpha^{2}}\frac{1}{y^{2}}
+\frac{2\beta^{2}}{\alpha^{2}}\frac{1}{y}-\frac{\beta^{2}}{\alpha^{2}}\right]R_{n}(y)=0
\end{equation}
In order to solve this equation with AIM for $\ell=0$, we should
transform this equation to the form of equation \ref{diff}.
Therefore, the reasonable physical wave function we propose is as
follows
\begin{equation}\label{wavem}
R_{n}(y)=y^{\frac{\varepsilon}{\alpha}}e^{-\frac{\beta}{\alpha}y}f_{n}(y)
\end{equation}
If we insert this wave function into the equation \ref{transm}, we
have the second-order homogeneous linear differential equations in
the following form
\begin{equation}\label{aimschrm}
\frac{d^{2}f_{n}(y)}{dy^{2}}=\left(\frac{2\beta y-2\varepsilon-\alpha}{\alpha y}\right)\frac{df_{n}(y)}{dy}
+\left(\frac{2\varepsilon\beta+\alpha\beta-2\beta^{2}}{y\alpha^{2}}\right)f_{n}(y)
\end{equation}
which is now amenable to an AIM solution. By comparing this equation
with equation \ref{diff}, we can write the $\lambda_{0}(y)$ and
$s_{0}(y)$ values and by means of equation \ref{iter}, we may
calculate $\lambda_k(y)$ and $s_k(y)$. This gives (the subscripts
are omitted):
\begin{eqnarray}\label{inim}
    \lambda_{0}&=&\left(\frac{2\beta y-2\varepsilon-\alpha}{\alpha y}\right)                   \nonumber \\
    s_{0}&=&  \left(\frac{2\varepsilon\beta+\alpha\beta-2\beta^{2}}{y\alpha^{2}}\right)        \nonumber \\
    \lambda_{1}&=& {\frac {-3\,\beta\,\alpha\,y+6\,\varepsilon\,\alpha+2\,{\alpha}^{2}-6\,
\beta\,y\varepsilon-2\,{\beta}^{2}y+4\,{\varepsilon}^{2}+4\,{\beta}^{2}{y}^{2}}{{\alpha}^{2}{y}^{2}}}  \nonumber \\
    s_{1}&=& 2\,{\frac {\beta\, \left( 2\,\varepsilon+\alpha-2\,\beta \right)
    \left(-\alpha-\varepsilon+\beta\,y \right) }{{\alpha}^{3}{y}^{2}}}
                                     \\ \ldots \emph{etc} \nonumber
\end{eqnarray}
Combining these results with the quantization condition given by
equation \ref{quantization} yields
\begin{eqnarray}
 \frac{s_0 }{\lambda _0 } & = & \frac{s_1 }{\lambda _1 }\,\,\,\,\,\, \Rightarrow
\,\,\,\,\,\,\varepsilon_{0}  =  -\frac{\alpha}{2}+\beta      \nonumber \\
 \frac{s_1 }{\lambda _1 }  & = & \frac{s_2 }{\lambda _2 }\,\,\,\,\,\, \Rightarrow
\,\,\,\,\,\, \varepsilon_{1}=   -\frac{3\alpha}{2}+\beta                            \nonumber \\
 \frac{s_2 }{\lambda _2 }  & = & \frac{s_3 }{\lambda _3 }\,\,\,\,\,\, \Rightarrow
\,\,\,\,\,\,\varepsilon_{2}=  -\frac{5\alpha}{2}+\beta                           \\
\ldots \emph{etc} \nonumber
 \end{eqnarray}
When the above expressions are generalized, the eigenvalues turn out
as
\begin{equation}\label{energym}
\varepsilon_{n}=\beta-(n+\frac{1}{2})\alpha
\hspace{1cm} n=0,1,2,3,...
\end{equation}
Using equation \ref{ansatzm}, we obtain the energy eigenvalues
E$_{n}$,
\begin{equation}\label{energyeigenm}
    E_{n}=-\frac{\hbar^{2}}{2\mu r_{e}^{2}}\left[\beta-(n+\frac{1}{2})\alpha\right]^{2}
\end{equation}
we calculate the energy eigenvalues of the $H_{2}$ diatomic molecule.
The AIM results are compared with those obtained by the hypervirial perturbation
method (HV) \cite{killingbeck}, the shifted 1/N and modified shifted 1/N
expansion methods \cite{bag} for the $H_{2}$ diatomic molecule in
Table \ref{Table2}. As it can be seen from the results presented
in these tables that the AIM results are in good agreement with
the findings of the other methods.

Now, As indicated in Section \ref{aim}, we can determine
corresponding wave function by using equation \ref{efson}. When we
compare equation \ref{compare1} and equation \ref{aimschrm}, we find
$N=-1$, $a=\frac{\beta}{\alpha}$, and
$m=\frac{2\varepsilon-\alpha}{2\alpha}$. Therefore, we find
$\sigma=\frac{2\varepsilon}{\alpha}+1$. For $b\rightarrow0$ we can
take the limit in equation \ref{efson} using the limit relation
\begin{equation}\label{limit}
\lim\limits_{b \to 0 } {_{2}}F_{1}(-n,\frac{1}{b}+a;c;zb) ={_{1}}F_{1}(-n;c;z)
\end{equation}

Consequently, the solution of equation \ref{aimschrm} can easily find
\begin{equation}
f_{n}(y)=(-1)^{n}\frac{\Gamma(\frac{2\varepsilon_{n}}{\alpha}+n+1)}{\Gamma(\frac{2\varepsilon_{n}}{\alpha}+1)}
{_{1}}F_{1}(-n,\frac{2\varepsilon_{n}}{\alpha}+1;\frac{2\beta}{\alpha}y)
\end{equation}
Thus, we can write the total radial wave function as below,
\begin{equation}\label{radyalwavem}
R_{n}(y)=Ny^{\frac{\varepsilon_{n}}{\alpha}}e^{-\frac{\beta}{\alpha}y}
{_{1}}F_{1}(-n,\frac{2\varepsilon_{n}}{\alpha}+1;\frac{2\beta}{\alpha}y)
\end{equation}
where $N$ is normalization constant.

\section{Conclusion}
\label{conclude} We have shown an alternative method to obtain the
energy eigenvalues and corresponding eigenfunctions of the deformed
Hulth\'{e}n and the Morse potentials within the
framework of the asymptotic iteration method. We have calculated the
energy eigenvalues for Hulth\'{e}n potential with $Z=1$, $q=1$ and
several values of the screening parameter. The AIM results are
compared with the Nikiforov-Uvarov \cite{nu1s} and the shifted 1/N
expansion \cite{other} methods in Table \ref{Table1}. Furthermore, We
have calculated the energy eigenvalues for $H_{2}$ diatomic molecule and
compared the hypervirial perturbation method \cite{killingbeck}, the shifted 1/N
and modified shifted 1/N expansion methods \cite{bag} in Table \ref{Table2}.
As it can be seen from the results presented in these tables, the AIM results are
in good agreement with the findings of the other methods.

The advantage of the asymptotic iteration method is that it gives the
eigenvalues directly by transforming the radial Schr\"{o}dinger
equation  into a form of ${y}''$ =$ \lambda _0 (r){y}' + s_0 (r)y$.
The method presented in this study is a systematic one and it is very
efficient and practical. It is worth extending this method to the
solution of other interaction problems.

\begin{table}[tbp]
\begin{center}
\begin{tabular}{ccccccccccccccc}
\hline\hline$n$&  & $-E_{n} (1/N)$ \cite{other}&  & $-E_{n}
(NU)$\cite{nu1s}& &$-E_{n} (AIM)$& \\ \hline
  & &&  &$\delta=0.002$&  \\
1&  & 0.4990005&  & 0.4990005&  & 0.4990005\\
2&  & 0.1240020&  & 0.1240020&  & 0.1240020\\
3&  & 0.0545601&  & 0.0545601&  & 0.0545601\\
4&  & 0.0302580&  & 0.0302580&  & 0.0302580\\
5&  &          &  & 0.0012500&  & 0.0012500\\  \\ \hline

  & &&  &$\delta=0.01$&  \\
1&  &0.4950125&  & 0.4950125&  & 0.4950125\\
2&  &0.1200500&  & 0.1200500&  & 0.1200500\\
3&  &0.0506681&  & 0.0506681&  & 0.0506681\\
4&  &0.0264501&  & 0.0264500&  & 0.0264500\\
5&  &0.0153128&  & 0.0153125&  & 0.0153125\\  \\ \hline

  & &&  &$\delta=0.05$&  \\
1&  &  0.4753125&  & 0.4753125&  & 0.4753125\\
2&  &  0.1012503&  & 0.1012500&  & 0.1012500\\
3&  &  0.0333746&  & 0.0333681&  & 0.0333681\\
4&  &  0.0113035&  & 0.0112500&  & 0.0112500\\
5&  &           &  & 0.0028125&  & 0.0028125\\  \\ \hline

  & &&  &$\delta=0.2$&  \\
1&  &   0.4049962 &  & 0.4050000 &  & 0.4050000 \\
2&  &   0.0450856 &  & 0.0450000 &  & 0.0450000 \\
3&  &             &  & 0.0005556 &  & 0.0005556 \\
4&  &             &  & 0.0112500 &  & 0.0112500 \\
5&  &             &  & 0.0450000 &  & 0.0450000   \\ \hline\hline
\end{tabular}%
\end{center}
\caption{The comparison of the AIM results (present work) with the
findings of the $1/N$ \cite{other} and NU \cite{nu1s} methods for
the s-states energy eigenvalues of the Hulth\'{e}n potential for
several values of screening parameter $\delta$.} \label{Table1}
\end{table}

\begin{table}
\begin{center}
\begin{tabular}{ccccccccccc}     \hline     \hline
\mbox{$n$}& \mbox{$\ell$}&\mbox{AIM}& \mbox{HV} \cite{killingbeck}&\mbox{Modified Shifted 1/N} \cite{bag}&\mbox{Shifted 1/N} \cite{bag} \\
 0     &     0   & -4.47601  & -4.47601    & -4.4760   &  -4.4749       \\
 5     &     0   & -2.22052  &  -2.22051   & -2.2205   &  -2.2038       \\
 7     &     0   & -1.53744  & -1.53743    & -1.5374   &  -1.5168       \\
\hline\hline
\end{tabular}
\end{center}
\caption{For the $H_{2}$ diatomic molecule, the comparison of the
energy eigenvalues (in eV) obtained by using AIM with other methods
for different values of $n$. Potential parameters are
$D~=~4.7446eV$, $a=1.9425 (A^{0})^{-1}$, $r_{e}=0.7416A^{0}$,
${\hbar c}=1973.29 eV A^{0}$ and $\mu=0.50391$amu.} \label{Table2}
\end{table}
\end{document}